\documentclass[numreferences]{kluwer}
\usepackage{graphicx}

\begin{document}
\begin{article}

\begin{opening}
\title{How black holes turn cusps into cores}
\author{Marc \surname{Hemsendorf}\email{marchems@physics.rutgers.edu}}
\institute{Dept.~of Physics and Astronomy, Rutgers University, 136
  Frelinghuysen Rd., Piscataway, NJ 08854, USA}

\begin{abstract}
  Collapsing collisionless particle systems form gravitational bound
  halos with cuspy density profiles. Also hierarchical merging of
  these systems produce remnants with cuspy central density
  profiles. These results lead to the assumption of cuspy NFW
  \cite{Navarro:96} profiles for the density distribution in dark
  matter halos. However, observed rotation curves in disk galaxies
  suggest dark matter halos with isothermal core. The same kind of
  problem can be found for globular clusters, which show cores in
  their density profiles but should be cuspy if they where formed
  through a cold collapse.\cite{King:62,Jenkins:98} We are showing how small,
  massive, and compact objects can efficiently transform cuspy stellar
  systems into density profiles with an isothermal core.
\end{abstract}


\runningtitle{Black holes and cusps}
\runningauthor{M.~Hemsendorf}
\end{opening}


\section{Introduction}

\section{Numerical experiments}
\subsection{Initial conditions}
  For the realizations of cuspy systems we are choosing Dehnen models
  with isotropic velocity dispersion. The Dehnen family of models are
  characterized by a parameter $\gamma$ that measures the degree of
  central concentration.\cite{Dehnen:93}  The density profile is
  \begin{equation}
    \rho(r) = \frac{(3 - \gamma) M}{4 \pi} 
    \; \frac{a}{r^{\gamma} \left(r + a\right)^{4-\gamma}},
    \label{eq:dehnen-rho}
  \end{equation}
  $M$ being the total mass and $a$ being the scale length. Both
  $M$ and $a$ are set to unity. We chose centrally condensed Dehnen
  models with inner slope $\gamma = 1$, $3/2$, and $2$, yielding a
  density profile similar to those of elliptical galaxies with dense
  stellar nuclei. The black holes and the stars are initially
  distributed according to the same profile. 

  We are creating several realizations of these Dehnen models with a
  total particle number of 9910. These realizations contain 10 black
  holes with a combined mass of 0.01 which is 1\% of the total mass
  of the system.

\subsection{Method}
  We are using NBODY6++ for the orbit integration. NBODY6++ is a high
  precision fourth order direct force
  integrator \cite{Spurzem:96,Aarseth:99-b}. Since it can regularize
  the equations of motion for particles having close encounters no
  softening needs to be applied. This is why we are able to follow the
  formation of black hole -- black hole binaries transfering part of
  their binding energy into the stellar system.

  We stopped the simulations once the number of black holes bound to
  the stellar system dropped below three. This process usually took a
  few hundred dynamical times. 

\subsection{Results}

  \begin{figure}
    \includegraphics[width=\linewidth]{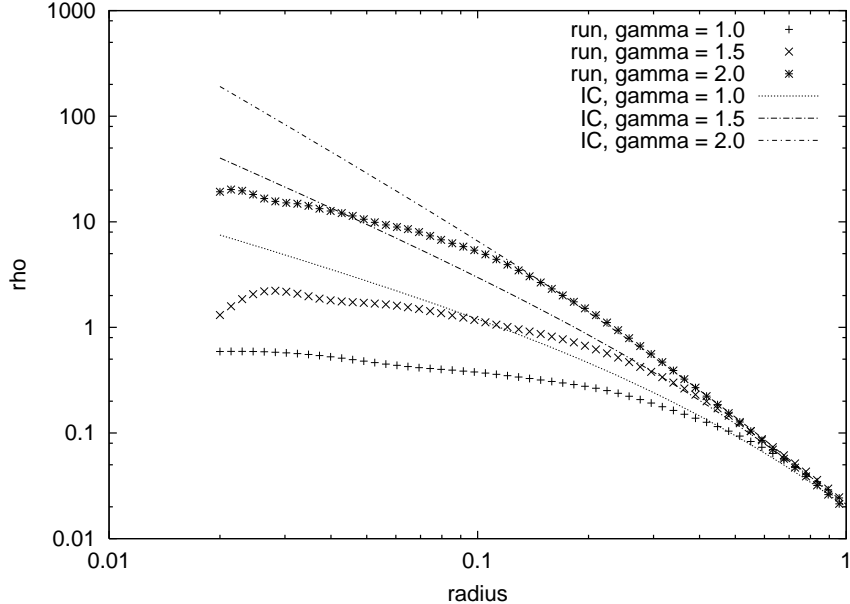}
    \caption{Comparison of the initial radial density distribution of
    the stars with the average radial density distribution of the
    stars after 600 $N$-body time units.}
    \label{fig:rho}
  \end{figure}

  \begin{figure}
    \includegraphics[width=\linewidth]{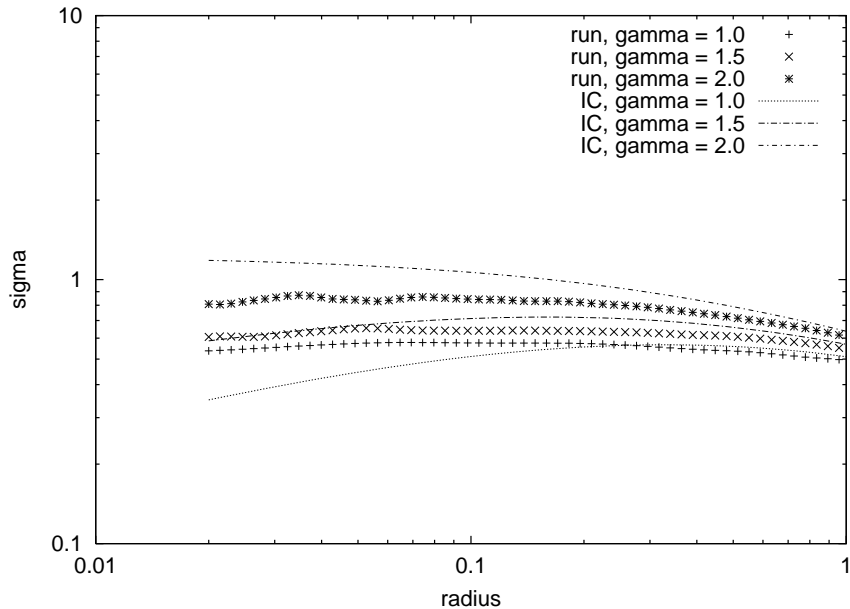}
    \caption{Comparison of the initial 3D velocity dispersion of the
    stars with the average 3D velocity dispersion of the stars after
    600 $N$-body time units.}
    \label{fig:sig}
  \end{figure}

  \begin{table}
    \caption{The amount of relocated mass after 600 $N$-body time units}
    \begin{tabular*}{12pc}{c|cc}
      $\gamma$ & $r_\mathrm{crit}$ & $M_\mathrm{rel}$ \\ \hline
      1 & 0.5 & 0.06 \\
      3/2 & 0.4 & 0.07 \\
      2 & 0.3 & 0.07
    \end{tabular*}
    \label{tab:mrel}
  \end{table}

  Figures \ref{fig:rho} and \ref{fig:sig} summarize our results. Both
  plots compare the initial model with the situation after 600
  $N$-body time units. This corrosponds to an evolution over roughly
  100 dynamical timescales. We have averaged the results of four
  realizations in order to find $\rho(r)$ and $\sigma(r)$ in Figures
  \ref{fig:rho} and \ref{fig:sig}.

  Figure \ref{fig:rho} shows the initial radial density distribution
  compared with the averaged radial density distribtion after 600
  $N$-body time units. While the impact of the black holes removes
  mass up to radius $r_\mathrm{crit}$ of 0.5 which is half way to the
  initial scale radius for $\gamma = 1.0$ and $\gamma = 1.5$, the
  effect is seemingly smaller in the models with $\gamma = 2.0$. As
  Table \ref{tab:mrel} shows, the amount of mass relocated by the
  black holes remains constant.

  Figure \ref{fig:sig} shows the initial 3D velocity dispersion in the
  models and the 3D velocity dispersion after 600 $N$-body time
  units. Since models with $\gamma = 2.0$ are isothermal, the
  evolution is quite modest for such models. In the case of smaller
  $\gamma$, Dehnen models have a zero velocity dispersion in the
  center.  Our models with smaller $\gamma$ show a strong evolution
  into isothermal systems through the influence of the black holes.

\section{Discussion}
  We have shown that 10 black holes are capable of efficiently
  destroying a stellar cusp within a few ten dynamical timescales. In
  our simulations each black hole removes an equivalent of seven times
  its own mass from the central regions of the sytem. The resulting
  systems show isothermal cores.

\begin{acknowledgements}
  This work was supported by NSF grant 00-71099 and by NASA grants
  NAG5-6037 and NAG5-9046. I would like to thank D.~Merritt and
  R.~Jimenez for their encouragement and help.  I am grateful to the
  NASA Center for Computational Sciences at NASA-Goddard Space Flight
  Center, the National Computational Science Alliance, the Center for
  Advanced Information Processing at Rutgers University, the John von
  Neumann Institut f\"ur Computing in J\"u\-lich, and the
  H\"ochst\-lei\-stungs\-re\-chen\-zen\-trum in Stuttgart for their
  generous allocations of computer time.
\end{acknowledgements}

\end{article}
\end{document}